\documentclass[12pt,a4paper]{article}

\pdfoutput=1

\usepackage[DIV13,BCOR0mm]{typearea}

\usepackage{graphicx}

\begin{document}

\title{\textbf{Numerical simulation of the evolution of
               glacial valley cross sections}}

\author{\textsc{Hakime Seddik}\thanks{E-mail: hakime@lowtem.hokudai.ac.jp}\\
        \textsc{Ralf Greve}\\
        \textsc{Shin Sugiyama}\\[0.5ex]
        {\normalsize Institute of Low Temperature Science, Hokkaido University,}\\[-0.25ex]
        {\normalsize Kita-19, Nishi-8, Kita-ku, Sapporo 060-0819, Japan}\\[1.5ex]
        \textsc{Renji Naruse}\\[0.5ex]
        {\normalsize Glacier and Cryospheric Environment Research Laboratory,}\\[-0.25ex]
        {\normalsize Higashi-machi 2-339, Tottori 680-0011, Japan}}

\date{}

\maketitle

\begin{abstract}

A numerical model was developed for simulating the formation of
U-shaped glacial valleys by coupling a two-dimensional ice flow
model with an erosion model for a transverse cross section. The
erosion model assumes that the erosion rate varies
quadratically with sliding speed. We compare the
two-dimensional model with a simple shallow-ice approximation
model and show the differences in the evolution of a
pre-glacial V-shaped valley profile using the two models. We
determine the specific role of the lateral shear stresses
acting on the glacier side walls in the formation of glacial
valleys. By comparing the model results with field data, we
find that U-shaped valleys can be formed within
$50\;\mathrm{ka}$. A shortcoming of the model is that it
primarily simulates the formation of glacial valleys by
deepening, whereas observed valleys apparently have formed
mainly by widening.

\end{abstract}

\section{Introduction}

Despite the fact that U-shaped valleys are characteristic
products of alpine glaciation, the interaction of ice flow and
glacial erosion which creates such well-known glacial forms has
not been widely studied. Empirical studies have established the
general concept that many glaciated valleys have approximately
parabolic (U-shaped) cross sections (Graf, 1970; Doornkamp and
King, 1971; Girard, 1976; Aniya and Welch, 1981). Both the
development of numerical ice-flow models (Reynaud, 1973; Budd
and Jensen, 1975; Mahaffy, 1976; Hook and others, 1979; Iken,
1981; Bindschadler 1982; Oerlemans, 1984) and the theoretical
and empirical work in geomorphology (Hallet, 1979, 1981;
Shoemaker, 1988, Iverson, 1990, 1991) have improved the
capability to simulate the characteristics of glacier motion
and the understanding of glacial erosion processes at small
scales.

While previous valley evolution studies from Harbor (1990,
1992, 1995) and MacGregor and others (2000) have provided a
better geomorphological understanding of glacial valley
formation, little has been done to investigate the
glaciological processes involved in their formation. The
glacial cross section evolution model used by Harbor (1990,
1992) successfully simulated a proper erosion pattern (central
minimum in the basal sliding velocity) for the U-shaped channel
development by assuming a quadratic function of the sliding
velocity for the erosion law, and it has identified the drag
process associated with it. However, Harbor (1990, 1992)
provided little information on the computation of the basal
shear stress, which is required for successful erosion
modeling. In particular, the detailed stress conditions
required for the development of a U-shaped valley have not been
described. Our work is complementary to the Harbor study and
describes how the lateral shear stress affects the development
of glacial valleys. 		

In this paper, we present the details of the two-dimensional
flow pattern computation and its coupling with the subglacial
erosion model. The study focuses in particular on the
investigation of the influence of the lateral shear stress
component on the formation of a U-shaped valley. For this
purpose, the two-dimensional model is compared with a simple
shallow-ice model in order to highlight the glacier stress
conditions favorable to the formation of U-shaped valleys.
Moreover, an attempt is made to constrain the basal sliding
parameter by comparing the model results with glacial valley
field data.

\section{Methods}

We developed a two-dimensional ice flow model in a transverse
section and coupled it to an erosion model. The flow model
solves the velocity field in a transverse cross section of a
glacier assuming a uniform geometry along the glacier. We used
a Cartesian coordinate with the $x$-axis along the glacier, $y$
across the glacier, and $z$ perpendicular to the $x$-$y$ plane
pointing upward (Fig.~1a). The stress components are
$\tau_{xx}$, $\tau_{yy}$, $\tau_{zz}$, $\tau_{xy}$,
$\tau_{xz}$, $\tau_{yz}$. All elements move along lines
parallel to the $x$-axis, so that the only velocity component
is $u$. We are interested in how $u$ varies with $y$ and $z$
and so assume that it does not depend on $x$. These assumptions
imply that the strain rates $\dot{\varepsilon}_{x}$,
$\dot{\varepsilon}_{y}$, $\dot{\varepsilon}_{z}$,
$\dot{\varepsilon}_{yz}$ are all zero. It follows that the
stress-deviator components $\tau_{xx}^\mathrm{D}$,
$\tau_{yy}^\mathrm{D}$, $\tau_{zz}^\mathrm{D}$,
$\tau_{yz}^\mathrm{D}$ are all zero, and the equilibrium
equation for the momentum balance of the ice in the $x$
direction is (Nye, 1965)
\begin{equation}
  \frac{\partial \tau_{xy}} {\partial y}
  + \frac{\partial \tau_{xz}} {\partial z}
  = -\rho g \sin{\alpha},
\end{equation}
where $\tau_{xy}$ and $\tau_{xz}$ are the shear stresses,
$\rho$ is the density of ice, $g$ the acceleration due to
gravity, and $\alpha$ the inclination angle of the glacier
surface (Fig.~1b). The $z$-component of the momentum balance
yields a hydrostatic distribution of the pressure $p$. Glacial
flow is treated as a non-Newtonian fluid, and Glen's flow law
is used as a constitutive relation with the viscosity $\mu$, so
that
\begin{equation}
  \frac{\partial u}{\partial y}
  = \frac{1}{\mu} \tau_{xy}, \ \ \ \frac{\partial u}{\partial z}
  = \frac{1}{\mu} \tau_{xz}.
\end{equation}
Equation~(2) can be written as
\begin{equation}
  \frac{\partial u}{\partial y} = 2F \tau_{xy}, \ \ \
  \frac{\partial u}{\partial z} = 2F \tau_{xz}.
\end{equation}
The term $F$ is the fluidity (here defined as one half of the
inverse viscosity), which can be factorized as
\begin{equation}
  F = A(\tau_\mathrm{e}^{2}+\tau_{0}^{2})^{\frac{n-1}{2}},
\end{equation}
where $\tau_\mathrm{e}$ is the effective stress, and the rate
factor $A$ and the flow-law exponent $n$ are material
parameters. We used the common values of $n=3$ and
$A=214\;\mathrm{MPa^{-3}\,a^{-1}}$ (Paterson, 1994). The
quantity $\tau_{0}$ is introduced to avoid the mathematical
singularity caused by an infinite viscosity when stresses
approach zero (Blatter, 1995).

\begin{figure}[htb]
  \centering
  \includegraphics[width=100mm]{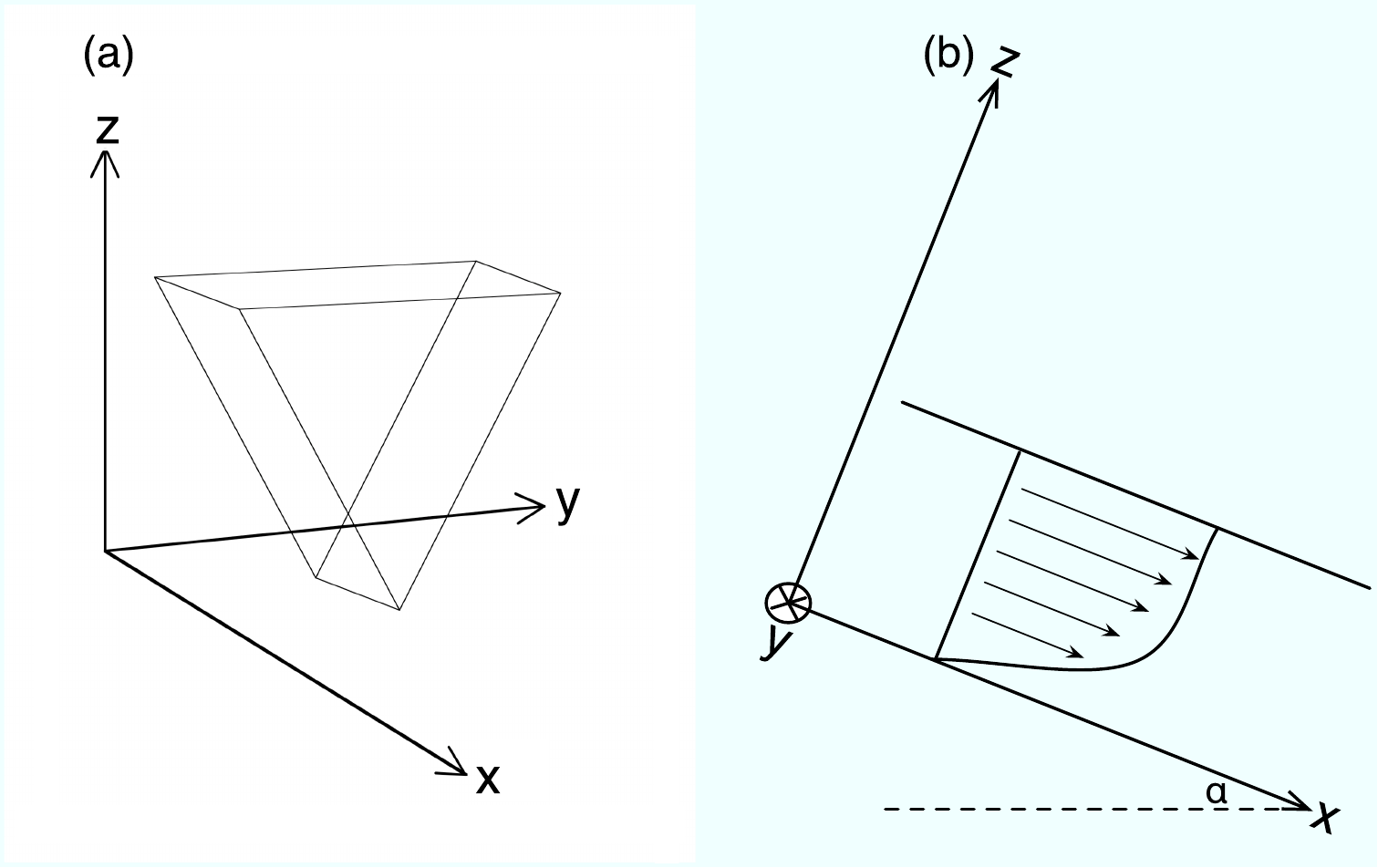}
  \caption{(a)~Initial valley geometry and coordinate system
  used in the simulations. (b)~Longitudinal glacier profile.}
\end{figure}

The boundary condition on the free surface $(z=S)$ consists of
vanishing pressure and shear traction,
\begin{equation}
  p_{s} = 0, \ \ \ \ \ \ \tau_{s} = \tau_{xz}=0.
\end{equation}
At the glacier base $(z=B(y))$, we introduce the basal sliding
by linearly relating the sliding speed $u_\mathrm{b}$ to the
shear stress acting on the bed $\tau_\mathrm{b}$ (Weertman,
1964; Lliboutry, 1968, 1979),
\begin{equation}
  u_\mathrm{b} = -c \ \tau_\mathrm{b},
\end{equation}
and
\begin{equation}
  \tau_\mathrm{b} = n_{y}\tau_{xy} + n_{z}\tau_{xz},
\end{equation}
\[
  n_{y} = \frac{1}{\sqrt{1+(\frac{\partial B}{\partial y})^{2}}}
          \frac{\partial B}{\partial y}, \ \ \ \
  n_{z} = -\frac{1}{\sqrt{1+(\frac{\partial B}{\partial y})^{2}}},
\]
where $c=50\;\mathrm{m\,a^{-1}\,MPa^{-1}}$ is our standard
value of the sliding coefficient, which is constant across the
glacier bed and has been chosen to obtain values for sliding
velocity that allow for glacial erosion. The flow speed at the
side margins is constrained to be zero.

The ice flow model was coupled with an erosion model by
introducing a quadratic function of the sliding speed for the
calculation of the erosion rate. Although the complex nature of
the glacial erosion has not allowed the development of
physically complete models for processes such as glacial
abrasion, plucking, subglacial fluvial erosion, and chemical
dissolution by subglacial water, this assumption for the
erosion law represents the general form of the abrasion law
proposed by Hallet (1979). Therefore, in the simulation
described here, erosion rate normal to the bedrock surface was
calculated as
\begin{equation}
  E = C\,u_\mathrm{b}^{2},
\end{equation}
where $C$ is an erosion constant equal to
$10^{-4}\;\mathrm{a\,m^{-1}}$ (Harbor, 1992; MacGregor and
others, 2000).	

\section{Numerical procedure}

We prescribe a V-shaped cross section with maximum ice
thickness of $480\;\mathrm{m}$, surface width of
$1200\;\mathrm{m}$, and downglacier slope of $4^{\circ}$
(Fig.~2a) as the initial glacier and valley geometries. We
employ a two-dimensional finite-difference grid with
$34\times{}35$ points to solve Eq.~(1) for the flow speed
within this cross section. The model solves a set of
finite-difference equations with the LU factorization method
assuming that the fluidity is constant and that the sliding
speed is zero. To solve numerically Eq.~(4) for the fluidity, a
Newton-Raphson scheme is employed, and the computed velocity
field is used so that the new values of the fluidity are
utilized in the next iteration step. The velocity field is also
utilized to compute the stress field with Eq.~(3) to introduce
sliding speed in the next step using Eq.~(6). The computation
is iterated until the velocity field converges within
$2\times{}10^{-4}\;\mathrm{m\,a^{-1}}$. 	

Coupling the ice-flow model and the erosion model allows for
investigating the temporal evolution of glacial valleys. For
the first time step, the above procedures are used to calculate
a flow pattern for the initial V-shaped valley. With the
calculated basal sliding speeds, Eq.~(8) is used to compute the
pattern of erosion rate across the profile. Then the new
coordinates for the glacier and the valley cross section are
calculated and used for the next time step. The ice surface
elevation is recalculated for each time-step by assuming a
constant cross-sectional ice area. Consequently, in order to
exclude ice-free points in the valley profile that emerge
during the simulation due to the lowering of the ice surface,
we allow the horizontal grid to shrink when the new
finite-difference grid is regenerated for the new cross
section. These procedures are repeated for a given number of
time steps. 	

For all simulations carried out in this study, the simulation
time scale is $50\;\mathrm{ka}$ with a time step of
$1\;\mathrm{ka}$, and it is assumed that there is no climate
forcing that constrains the growth or the decline of the
glacier.

\section{Results and discussion}

\subsection{Full two-dimensional model vs.\ shallow-ice model}

As stated in the introduction, the first motivation of this
work is to investigate the glaciological conditions favorable
to the development of U-shaped valleys. We are interested in
the investigation of the stress conditions that are required
for the development of glacial valleys, and more precisely for
the gradual transformation of the V-shaped pre-glacial valley
to a U-shaped profile.

\begin{figure}[p]
  \centering
  \includegraphics[width=0.95\textwidth]{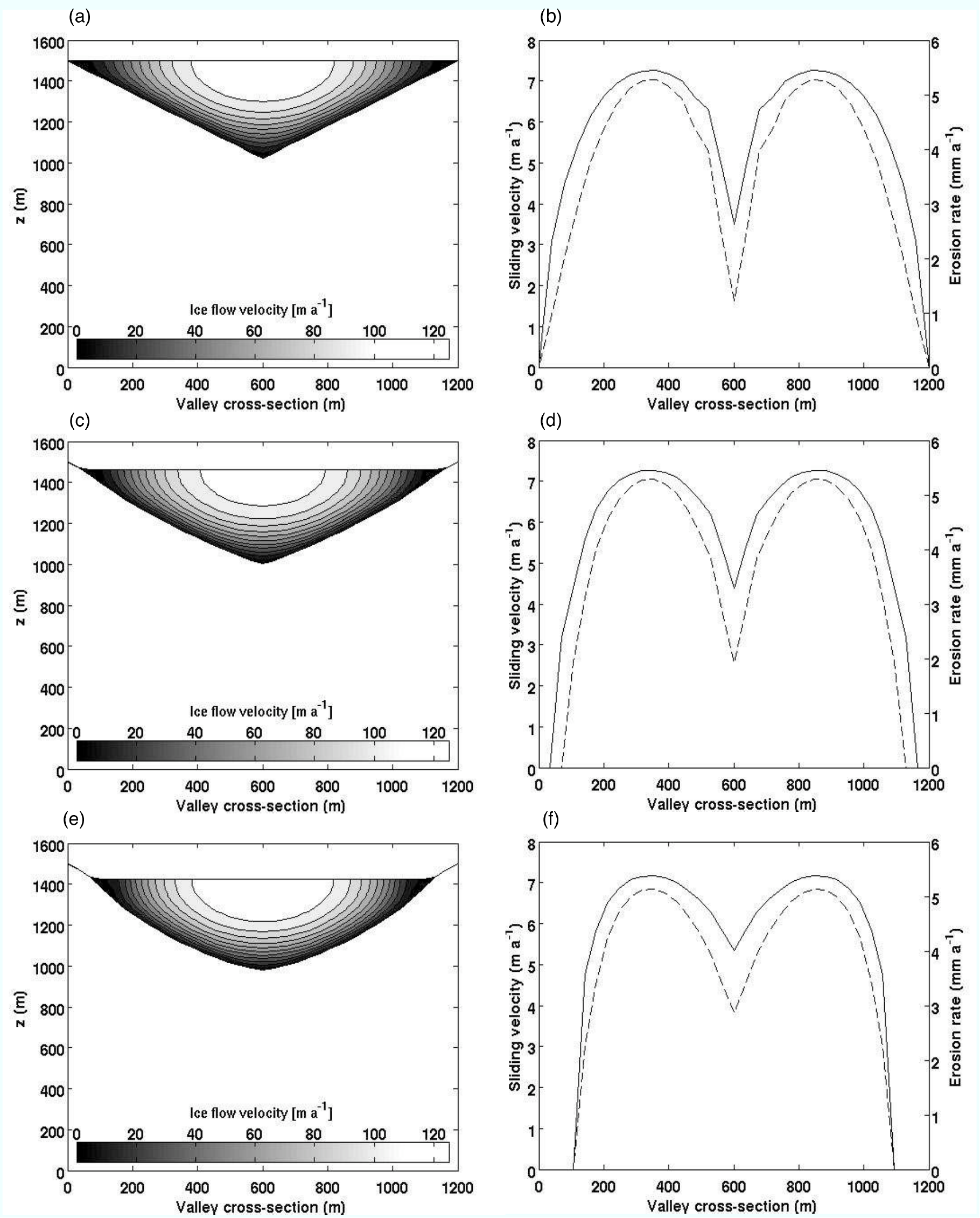}
  \caption{(a), (c), (e), (g), (i), (k)~Glacial valley evolution from
  an initial pre-glacial profile computed with the
  two-dimensional model for times $t=0,$ 10, 20, 30, 40
  and $50\;\mathrm{ka}$.
  (b), (d), (f), (h), (j), (l)~Corresponding cross-glacier variation
  of sliding velocities (solid lines) and erosion rates (dashed lines).}
\end{figure}

\addtocounter{figure}{-1}

\begin{figure}[p]
  \centering
  \includegraphics[width=0.95\textwidth]{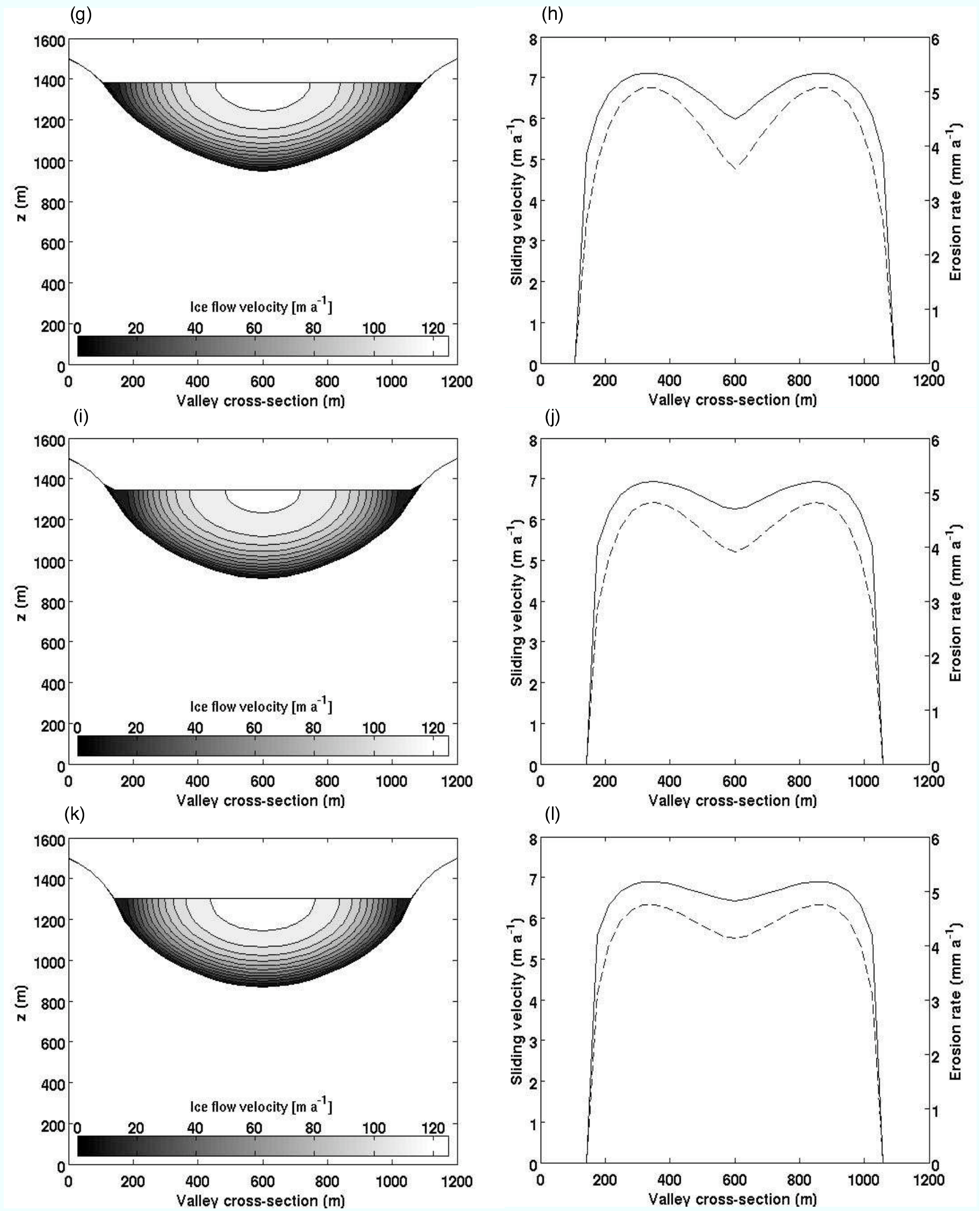}
  \caption{[cont.] (a), (c), (e), (g), (i), (k)~Glacial valley evolution from
  an initial pre-glacial profile computed with the
  two-dimensional model for times $t=0,$ 10, 20, 30, 40
  and $50\;\mathrm{ka}$.
  (b), (d), (f), (h), (j), (l)~Corresponding cross-glacier
  variation of sliding velocities (solid lines) and erosion rates
  (dashed lines).}
\end{figure}

Figures~2a and 2b show the two-dimensional velocity field
computed with the model described in Sect.~2 and the
cross-sectional sliding velocities and erosion rates (for the
initial V-shaped valley). Characteristic features of the
sliding velocity and the erosion rate are the increase toward
the interior of the glacier, but with local minima at the
center of the cross profile. The increase of drag (whose origin
is discussed later in this section) associated with the tight
form at the center of the V-shaped valley reduces velocities
there (Fig.~2b), resulting in a central minimum of the erosion
(required to convert a V-shaped valley to a U-shaped channel).
Figures~2a-l show the complete development sequence of a
pre-glacial V-shaped valley to a recognizable U-shaped profile
over $50\;\mathrm{ka}$, and the associated sliding velocities
and erosion rates every $10\;\mathrm{ka}$. The model predicts
the evolution of the V-shaped profile into a recognizable
glacial form with sliding velocities ranging from
$6\;\mathrm{m\,a^{-1}}$ to $7\;\mathrm{m\,a^{-1}}$, and for the
last time-step, the model could simulate the formation of a
deep and well-defined U-shaped valley (Fig.~2k).

\begin{figure}[htb]
  \centering
  \includegraphics[width=0.95\textwidth]{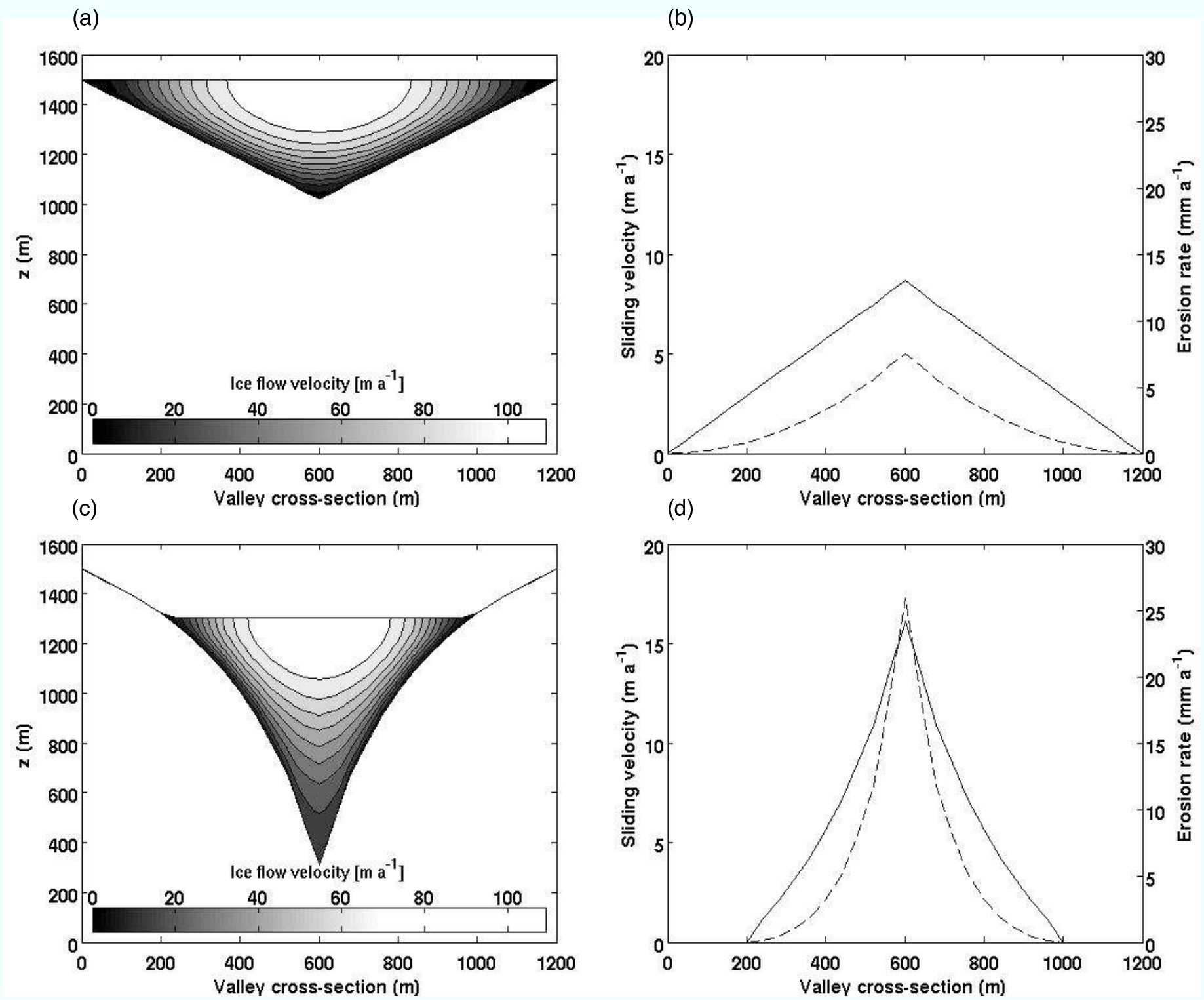}
  \caption{(a), (c)~Initial V-shaped valley and eroded valley
  obtained with the shallow-ice model.
  (b), (d)~Corresponding
  cross-glacier variation of sliding velocities (solid line) and
  erosion rates (dashed line).}
\end{figure}	

Although the time scale is sensitively dependent on the value
of the erosion constant in the erosion law, the model suggests
that a glacial valley can be developed after $50\;\mathrm{ka}$
or during a single glaciation under the condition of realistic
sliding speed. This observation agrees with previous results
from Harbor (1992). As the valley is progressively transformed
into a U-shaped form, the central minima in velocity and
erosion rates gradually decrease and practically disappear
during the last time-step (Fig.~2l). 	

For comparison with the full two-dimensional model, we now run
a shallow-ice model that does not compute the lateral shear
stress. The shallow-ice model only computes the shear stress
parallel to the bed, therefore omitting the first term in
Eq.~(1), and by integration of the simplified equation, the
basal shear stress is
\begin{equation}
  \tau_\mathrm{b} = -\tau_{xz}(z=B)=-\rho g H \sin{\alpha},
\end{equation}
where $H$ is the ice thickness. We run the shallow-ice model
with the same simulation procedure and initial setting as the
full two-dimensional model. As shown in Fig.~3b, the pattern of
the sliding velocity for the initial V-shaped profile is
directly proportional to the ice thickness. Due to the nature
of the erosion law, erosion values follow the same pattern.
This pattern of erosion was applied to the initial V-shaped
valley and Fig.~3c shows the glacial valley eroded during
$50\;\mathrm{ka}$. As observed, the valley shape remains
similar to a V-shaped profile, characterized by a deep and
narrow channel at the valley center. Clearly the ice thickness
dependent basal shear stress in the shallow ice model results
in higher velocities and erosions rates at the valley center
because this stress configuration ignores the side walls effect
on the ice flow and therefore prevents the development of a
U-shaped profile. Harbor (1992) and our full two-dimensional
model have clearly identified the friction effect that
decreases the sliding velocity at the center of the V-shaped
profile. In glaciological terms, the minimum basal velocity at
the valley center is the consequence of  the high drag between
the narrow side walls, which reduces  the deformation by shear
at this location. The associated minimum erosion at the profile
center leads to the gradual transformation of the V-shaped
profile into a U-shaped form, the maximum of erosion being
transferred closer to the valley margins. This process allows
the profile to widen toward a glacial valley shape. In the case
of the shallow-ice model, such a process did not appear
(Fig.~3d) and the resulting eroded valley still exhibits a
V-shaped form.

The shallow-ice model allows us to compare the results computed
with the full two-dimensional flow model in order to identify
the primary factor of the drag effect described by Harbor
(1990, 1992). This comparison shows clearly the importance of
the lateral shear from the side walls, i.e., $\tau_{xy}$, for
the modeling of glacial erosion in a valley cross section.
Ultimately, inclusion of both shear stress components
$\tau_{xy}$ and $\tau_{xz}$ is necessary to obtain the sliding
velocity pattern required for U-shaped valley formation.

\subsection{Comparison with field data}

Together with our motivation for the investigation of the
stress conditions for glacial valley development, comparison
with field data is needed for model validation. To describe the
differences in morphology between valley profiles and in many
glacial valley studies (Svensson, 1959; Harbor, 1992; Hirano
and Aniya, 1988), it has been common to use a power law
equation as a mathematical function to represent the glacial
trough cross-profile. The power law function is written as
\begin{equation}
  z = ay^{b},
\end{equation}
where $y$ and $z$ are the horizontal and vertical distances
from the lowest point of the cross section, and $a$ and $b$ are
constants. The value $b$ is commonly used as an index of the
steepness of the valley side, and $a$ a measure of the breadth
of the valley floor. Previous studies suggested that the valley
morphology progressively approaches a true parabolic form with
increasing glacial erosion, and that stage of valley evolution
can thus be measured by the proximity of $b$ to 2 (Svensonn,
1959; Graf, 1970; Hirano and Aniya, 1988). For further
comparison, we also use a form ratio $F\!R$, calculated by
\begin{equation}
  F\!R = \frac{D}{W},
\end{equation}
where $D$ is the valley depth and $W$ is the valley top width.
A large value of $F\!R$ depicts then an overdeepening
development of the glacial valley.

Only a few data describing present glacial valley shapes are
available, and studies that try to determine the time scale for
their formation by field measurements and profile
reconstruction are still lacking. Consequently, our comparison
will focus on investigating the shape of the glacial valley
computed by the model and the shape measured in natural
valleys. This is particularly useful in order to examine the
sliding velocity values required to obtain a profile shape
closed to the measured one. For our comparison work, we use the
field measurements obtained by Yingkui and others (2001) in the
Tian Shan Mountains, because they provide a rich set of form
coefficient values calculated for several measured profiles. As
described in their study, the morphological characteristics of
glacial valley cross sections in the middle and western Tian
Shan Mountains is represented by large variations of power law
coefficients. Values of $b$ in these areas range from 1.027 to
3.503, with most values in range 1.3-2.5. The average $F\!R$
values of tributaries are commonly larger than those for the
main valleys in these areas. 	

For our comparison study, our focus is the calculation of the
exponent $b$ and the form ratio $F\!R$ values for the valley
profile computed by the model, therefore the calculation of the
value of $a$ in Eq.~(10) is ignored. We run the model with
several values of the sliding coefficient $c$ (from 10 to
$50\;\mathrm{m\,a^{-1}\,MPa^{-1}}$) used in the sliding law
[Eq.~(6)], in order to constrain the glacier sliding velocity.
To ensure the convergence of the velocity field during the
computation, we use several grid resolutions ranging from
$34\times{}35$ to $94\times{}95$ points. Figure~4a shows the
comparison between the $b$ values provided by Yingkui and
others (2001) for 48 profiles and those calculated with the
computed profiles using different sliding coefficients, i.e.,
different sliding velocity conditions. As observed, the $b$
values for the profiles computed by the model range from 1.13
to 2.05, and therefore they generally fit to the distribution
of values observed by Yingkui and others (2001) where most
values range from 1.3 to 2.5. More interestingly, the $b$
values from the model results that fit most the general trend
of the observed distribution are those where the sliding
coefficient values are between 30 and
$50\;\mathrm{m\,a^{-1}\,MPa^{-1}}$. Consequently, a glacier
with sliding velocities greater than 3-$4\;\mathrm{m\,a^{-1}}$
would already be able to form a glacial valley shape similar to
some observed valley profiles. Moreover, as it is usually
assumed that a value of 2 for the exponent $b$ indicates a
valley evolution that has reached a U-shape form, the model
suggests that such form is obtained with sliding velocities
greater than $6\;\mathrm{m\,a^{-1}}$. As mentioned earlier, all
the simulations have been run within $50\;\mathrm{ka}$ because
$50\;\mathrm{ka}$ is a reasonable time-scale for the formation
of the Tian Shan Mountains valleys during the last glacial
period, even though the study by Yingkui and others (2001) did
not provide a direct confirmation of this time scale. In that
sense, our model results show that the formation within
$50\;\mathrm{ka}$ of glacial valleys similar in shape to the
Tian Shan mountains valleys is possible.
		
\begin{figure}[htb]
  \centering
  \includegraphics[width=0.95\textwidth]{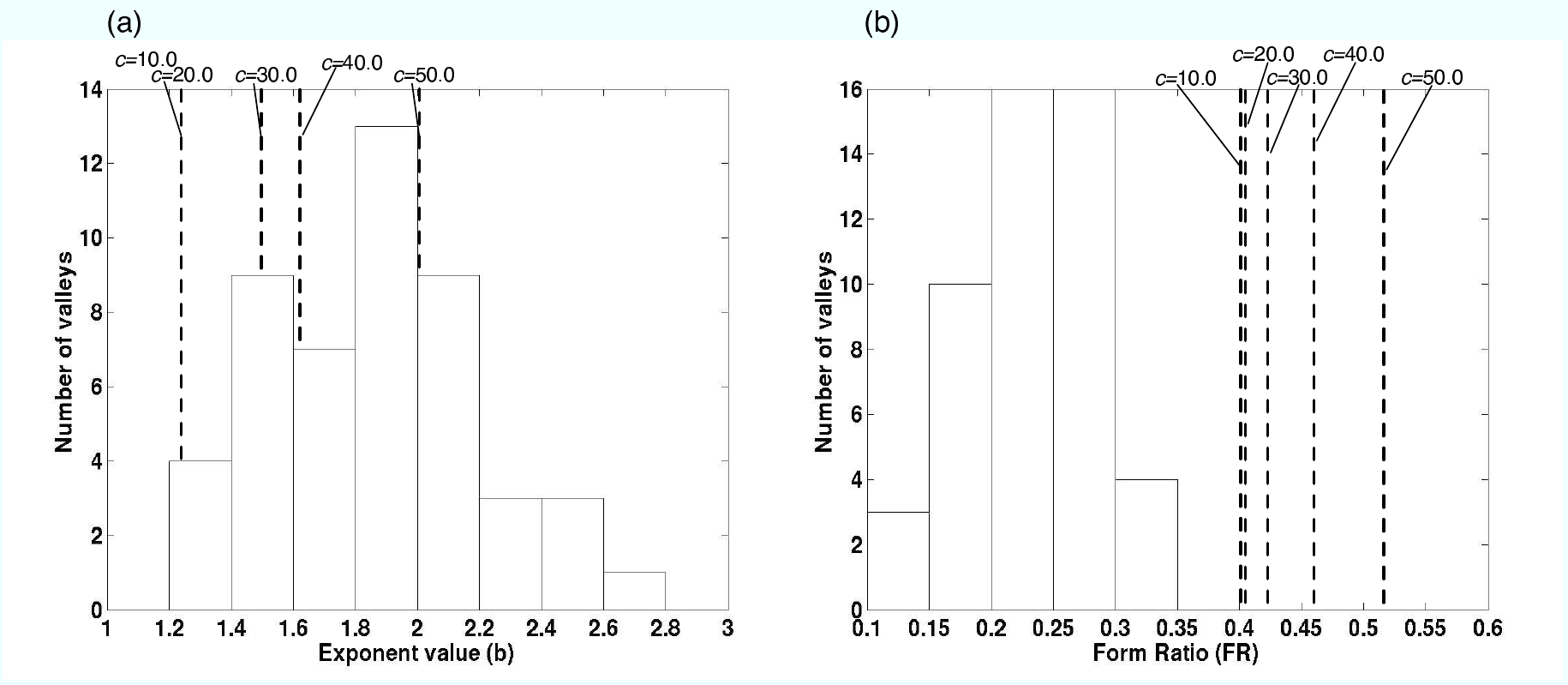}
  \caption{Comparison of the values of (a) the exponent $b$
  and (b) the form factor $F\!R$ obtained with the model (dashed
  lines), and corresponding values obtained by Yingkui and others (2001)
  for the Tian Shan Mountains (histogram).}
\end{figure}

\begin{figure}[htb]
  \centering
  \includegraphics[width=100mm]{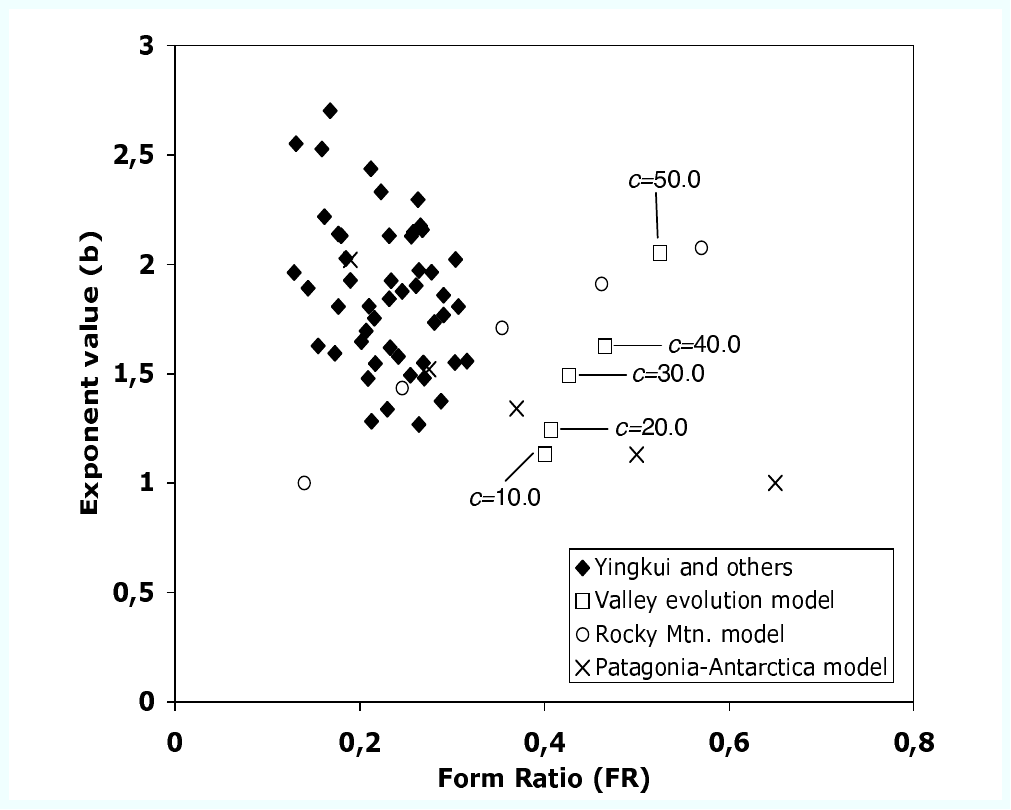}
  \caption{`$b$-$F\!R$ diagram' for the glacial valley
  cross sections measured by Yingkui and others (2001) and for
  the profiles computed by the model with sliding coefficients
  $c$ ranging from 10 to $50\;\mathrm{m\,a^{-1}\,MPa^{-1}}$.
  The values for the Rocky Mountain and Patagonia-Antarctica
  models (Hirano and Aniya, 1988) are plotted for comparison.}
\end{figure}

The comparison of the form ratio values obtained with the
profiles computed by the model and those provided by field
measurements is shown in Fig.~4b. We can clearly see that the
general distribution trend of the form ratios between the model
results and the measured profiles are significantly different.
While the maximum values of $F\!R$ provided by field data are
close to 0.3, the results obtained by the model gives higher
values ranging from 0.40 to 0.53. This is an important
divergence of the models results from field data, as it
indicates that the computed profile develops essentially by
deepening without widening as opposed to the widening without
deepening trend observed in the measured valleys. Better
description of the development process of glaciated valley
morphology can be addressed by describing the relationship
between $b$ values and form ratios within a `$b$-$F\!R$
diagram'. Hirano and Aniya (1988) have identified with their
model two opposite trends in glacial valley development by
analyzing the `$b$-$F\!R$ diagram' for several studied valleys.
The Rocky Mountain model depicts an overdeepening development
of the glacial valleys, and the Patagonia-Antarctica model
indicates a widening rather than deepening process of the
glacial valley development. Figure~5 shows the `$b$-$F\!R$
diagram' for the Tian Shan Mountains valleys and the model
results computed with the sliding coefficients used in Fig.~4.
The model results show larger $b$ values with increasing form
ratios (and increasing maximum velocities), as opposed to the
data for the Tian Shan Mountains, which indicate smaller $b$
values with increasing form ratios. For this reason, the
measured valleys correspond to the Patagonia-Antarctica model,
whereas the model results show the similar trend of the Rocky
Mountain model (Fig.~5). The difference in the glacial valley
development processes usually reflects the difference in the
initial V-shape of the valley, the initial relief, the ice
thickness, the lithology and structure of the local rock and
the glacial history. Although the V-shaped valley used as
initial condition in the simulation could also explain why the
computed profile only develops by deepening, the sliding
velocity based erosion law is the main reason of this
deepening. With increasing values of ice thickness, sliding
velocities are higher and the erosion tends to be more
important near the valley center, so that the valley deepening
is faster than the widening. Consequently, as our model is not
able to simulate the primary widening process observed in some
glacial valleys, future investigations of glacial valley
development modeling should focus on improving the erosion law.
However, a sliding velocity based erosion model seems to be
successful to simulate the Rocky Mountain model for the
development process of glaciated valley morphology identified
by Hirano and Aniya (1988). Running the model with a constant
ice surface altitude and increasing ice volume gives the same
pattern of valley development as previously described, i.e.,
development by deepening.

\begin{figure}[htb]
  \centering
  \includegraphics[width=0.95\textwidth]{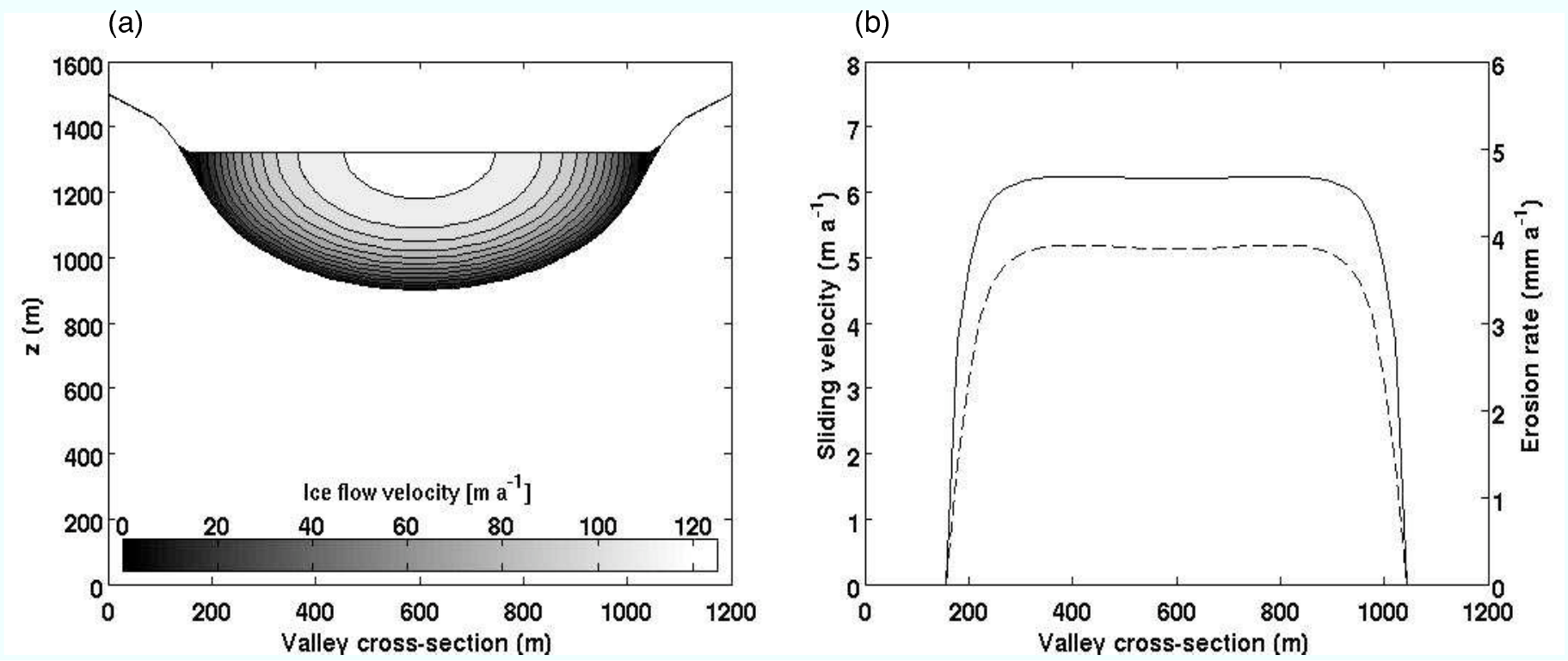}
  \caption{Model results obtained with the effective pressure
  included in the sliding law ($p=3$, $q=2$). (a) Glacial valley
  cross section at $50\;\mathrm{ka}$. (b) Corresponding cross-glacier
  variation of sliding velocities (solid line) and erosion rates
  (dashed line).}
\end{figure}

In order to investigate the possible influence of the sliding
law in the erosion and in the development of glacial valleys,
an alternative sliding law to the one described in Eq.~(6) may
be considered. For this purpose we use a sliding law which is
also dependent on the effective pressure, so that the basal
sliding is related to the basal shear stress $\tau_\mathrm{b}$
and the effective pressure $N$ by (Paterson, 1994)
\begin{equation}
  u_\mathrm{b} = k\frac{\tau_\mathrm{b}^{p}}{N^{q}},
\end{equation}
where $k$ is the sliding coefficient which is constant across
the glacier bed, and $p$ and $q$ are constants equal to 3 and
2, respectively. The value of $k$ is chosen in order to obtain
sliding velocities in the range of those obtained in Fig.~2l.
The effective pressure is defined by the ice overburden
pressure $p_\mathrm{i}$ and the water pressure $p_\mathrm{w}$
as $N=p_\mathrm{i}-p_\mathrm{w}$.

We neglect the water layer, so that the effective pressure
reduces to the ice overburden pressure. Figure~6 shows the
developed glacial valley and the corresponding sliding
velocities and erosion rates. Compared with Fig.~2l, the
developed profile exhibits similar U-shaped topography, however
calculated velocity and erosion values minima at the valley
center have totally disappeared, suggesting that the U-shaped
is more pronounced. This is confirmed by the $b$ value (2.19),
which is slightly higher than previously obtained in Fig.~4a
(2.05 for $c=50\;\mathrm{m\,a^{-1}\,MPa^{-1}}$). The $F\!R$
value (0.5) is in the same range as in Fig.~4b (0.53).
Therefore, the change in the sliding velocity law does not
affect significantly the overall results presented in this
paper.

\section{Conclusion}

This work presented a two-dimensional flow model of a glacier
in a valley cross section, which has been coupled with an
erosion law for the study of glacial valleys development. The
model has successfully described the stress conditions inside
the glacier required for the development of the glacial valley.
Comparison with a shallow-ice model showed the importance of
the lateral shear stress (lateral drag) in order to obtain the
proper pattern of the sliding velocity and the erosion in the
initial V-shaped valley for the development of the U-shaped
profile.

Comparison with field data obtained in the Tian Shan Mountains
allowed for constraining the value of the sliding coefficient
in the basal sliding law so that such forms, i.e., U-shaped
profiles, could be obtained within $50\;\mathrm{ka}$. However,
the process of formation of the Tian Shan Mountains valleys
could not be simulated with our model. Notably, different
values for the form ratio compared with the measured valleys
showed that the development of glacial valleys simulated with a
sliding velocity based erosion law occurs by a deepening
without widening process (Rocky Mountain model). However, the
measured valleys have developed with a widening without
deepening process (Patagonia-Antarctica model), which our model
was not able to simulate due to the way the erosion law was
defined.

Future improvements of the valley evolution model should focus
on more comparisons with field data and on the enhancement of
the erosion law. Improving the simulation of the erosion
occurring at the valley margins should provide a better way to
simulate valleys that correspond to the Patagonia-Antarctica
model. Further, a three-dimensional model would certainly
provide a better understanding of the evolution of the glacial
valley cross section together with the evolution of the
longitudinal profile.

\subsection*{Acknowledgements}

We wish to extend our thanks to Heinz Blatter (Institute for
Atmospheric and Climate Science, ETH Zurich) for providing
useful lecture notes on glacier modeling, and to Jean-Luc
Mercier (Faculty of Geography, Louis Pasteur University
Strasbourg) for his comments and interest in this work.
Furthermore, we are grateful to many colleagues at the
Institute of Low Temperature Science for helpful discussions.

\subsection*{References}

\begin{footnotesize}

\begin{description}
\frenchspacing

\item Aniya, M. and Welch, R., 1981. Morphological analyses
    of glacial valleys and estimates of sediment thickness
    on the valley floor: Victoria Valley system,
    Antarctica. \emph{The Antarctic Record}, \textbf{71},
    76--95.

\item Bindschadler, R., 1982. A numerical model of
    temperate glacier flow applied to the quiescent phase
    of a surge-type glacier. \emph{J.~Glaciol.},
    \textbf{28}, 239--265.

\item Blatter, H., 1995. Velocity and stress fields in
    grounded glaciers: a simple algorithm for including
    deviatoric stress gradients. \emph{J.~Glaciol.},
    \textbf{41}, 333--344.

\item Budd, W.F. and Jensen, D., 1975. Numerical modeling
    of glacier systems. International Association of
    Hydrological Sciences Publication No.~104, 257--291.

\item Doornkamp, J.C. and King, C.A.M., 1971.
    \emph{Numerical Analysis in Geomorphology}. Edward
    Arnold, London, England, 608~p.

\item Girard, W.W., 1976. Size, shape and symmetry of the
    cross-profiles of glacial valleys. Ph.D. thesis, Iowa
    City, Iowa, University of Iowa, 90~p.

\item Graf, W.L., 1970. The geomorphology of the glacial
    valley cross-section. \emph{Arctic and Alpine
    Research}, \textbf{2}, 303--312.

\item Hallet, B., 1979. A theoretical model of glacial
    abrasion. \emph{J.~Glaciol.}, \textbf{23}, 39--50.

\item Hallet, B., 1981. Glacial abrasion and sliding: Their
    dependence on the debris concentration in basal ice.
    \emph{Ann. Glaciol.}, \textbf{2}, 23--28.

\item Harbor, J.M., 1990. Numerical modeling of the
    development of glacial-valley cross sections. Ph.D.
    thesis, University of Washington, Seattle, Washington,
    219~p.

\item Harbor, J.M., 1992. Numerical modeling of the
    development of U-shaped valleys by glacial erosion.
    \emph{Geological Society of America Bulletin},
    \textbf{104}, 1364--1375.

\item Harbor, J.M., 1995. Development of glacial-valley
    cross-sections under conditions of spatially variable
    resistance to erosion. \emph{Geomorphology},
    \textbf{14}, 99--107.

\item Hirano, M. and Aniya, M., 1988. A rational
    explanation of cross-profile morphology for glacial
    valleys and of glacial valley development.  \emph{Earth
    Surface Processes and Landforms}, \textbf{13},
    707--716.

\item Hooke, R.L., Raymond, C.F., Hotchkiss, R.L. and
    Gustafson, R.J., 1979. Calculation of velocity and
    temperature in a polar glacier using the finite-element
    method. \emph{J.~Glaciol.}, \textbf{24}, 131--146.

\item Iken, A., 1981. The effect of the subglacial water
    pressure on the sliding velocity of a glacier in an
    idealized numerical model. \emph{J.~Glaciol.},
    \textbf{27}, 404--421.

\item Iverson, N.R., 1990. Laboratory simulations of
    glacial abrasion: Comparison with theory.
    \emph{J.~Glaciol.}, \textbf{36}, 304--314.

\item Iverson, N.R., 1991. Potential effects of subglacial
    water-pressure on quarrying. \emph{J.~Glaciol.},
    \textbf{37}, 27--36.

\item Lliboutry, L., 1968. General theory of subglacial
    cavitation and sliding of temperate glaciers.
    \emph{J.~Glaciol.}, \textbf{7}, 21--58.

\item Lliboutry, L., 1979. Local friction laws for
    glaciers: a critical review and new openings.
    \emph{J.~Glaciol.}, \textbf{23}, 67--95.

\item MacGregor, K.R., Anderson, R.S., Anderson, S.P. and
    Waddington, E.D., 2000. Numerical simulations of
    glacial-valley longitudinal profile evolution.
    \emph{Geol.}, \textbf{28}, 1031--1034.

\item Mahaffy, M.W., 1976. A three-dimensional numerical
    model of ice sheets. Test on the Barnes Ice Cap,
    Northwest Territories. \emph{J.~Geophys. Res.},
    \textbf{81}, 1059--1066.

\item Nye, J.F., 1965. The flow of a glacier in a channel
    of rectangular, elliptic or parabolic cross-section.
    \emph{J.~Glaciol.}, \textbf{5}, 661--690.

\item Oerlemans, J., 1984. Numerical experiments on
    large-scale glacial erosion. \emph{Zeit\-schrift f\"ur
    Glet\-scher\-kun\-de und Gla\-zial\-geo\-lo\-gie},
    \textbf{20}, 107--126.

\item Paterson, W.S.B., 1994. \emph{The Physics of
    Glaciers}. 3rd edition, Pergamon Press, Oxford etc.,
    480~p.

\item Reynaud, L., 1973. Flow of a valley glacier with a
    solid friction law. \emph{J.~Glaciol.}, \textbf{12},
    251--258.

\item Shoemaker, E.M., 1988. On the formulation of basal
    debris drag for the case of sparse debris.
    \emph{J.~Glaciol.}, \textbf{34}, 259--264.

\item Svenson, H., 1959. Is the cross-section of a glacial
    valley a parabola? \emph{J.~Glaciol.}, \textbf{3},
    362--363.

\item Yingkui, L., Gengnian, L. and Zhijiu, C., 2001.
    Glacial valley cross-profile morphology, Tian Shan
    Mountains, China. \emph{Geomorphology}, \textbf{38},
    153--166.

\item Weertman, J., 1964. The theory of glacier sliding.
    \emph{J.~Glaciol.}, \textbf{5}, 287--303.

\nonfrenchspacing
\end{description}

\end{footnotesize}

\end{document}